\documentclass[conference]{IEEEtran}
\IEEEoverridecommandlockouts

\usepackage{cite}
\usepackage{amsmath,amssymb,amsfonts}
\usepackage{graphicx}
\usepackage{textcomp}
\usepackage{xcolor}
\def\BibTeX{{\rm B\kern-.05em{\sc i\kern-.025em b}\kern-.08em
    T\kern-.1667em\lower.7ex\hbox{E}\kern-.125emX}}

\usepackage{caption}
\usepackage{subcaption}

\usepackage{graphicx}
\usepackage{float}

\usepackage{multirow}
\usepackage{booktabs}

\usepackage{xcolor}
\usepackage{soul}

\usepackage[linesnumbered,ruled]{algorithm2e}
\newcommand{\textcode}[1]{\textbf{\textup{#1}}}

\newcommand{\fun}[1]{\textit{\textbf{#1(}\textbf{)}}}
\newcommand{\funa}[2]{\textit{\textbf{#1(}$\mathtt{#2}$\textbf{)}}}
\newcommand{\funb}[3]{\textit{\textbf{#1(}$\mathtt{#2}$, $\mathtt{#3}$\textbf{)}}}
\newcommand{\func}[4]{\textit{{\textbf{#1(}$\mathtt{#2}$, $\mathtt{#3}$, $\mathtt{#4}$\textbf{)}}}}

\usepackage{pifont}

\usepackage{url}
\usepackage{hyperref}
\hypersetup{
    colorlinks=true,      
    linkcolor=blue,       
    citecolor=magenta,    
    filecolor=cyan,       
    urlcolor=black          
}

\usepackage{bbding}
\usepackage{multirow}

\usepackage{amsmath,amsthm,mathtools}

\usepackage[switch]{lineno}

\newcommand{\tab}[1]{TABLE~\ref{tab:#1}}
\newcommand{\fig}[1]{Fig.~\ref{fig:#1}}

\newcommand{\alg}[1]{Algorithm~\ref{alg:#1}}
\newcommand{\sect}[1]{\S\ref{sect:#1}}

\newcommand{\etal}{\textit{et al.}\xspace}

\newcommand{\name}{\texttt{AARC}\xspace}

\begin{document}

\title{AARC: Automated Affinity-aware Resource Configuration for Serverless Workflows}

\author{
    \IEEEauthorblockN{Lingxiao~Jin$^{1*}$, Zinuo~Cai$^{1*}$, Zebin~Chen$^{1}$, Hongyu~Zhao$^1$, Ruhui~Ma$^{1}$}
    \IEEEauthorblockA{$^1$School of Electronic Information and Electrical Engineering, Shanghai Jiao Tong University, Shanghai, China}
    \IEEEauthorblockA{\{jinlingxiao1122, kingczn1314, czb453874483, sjtu-zhy, ruhuima\}@sjtu.edu.cn}

\thanks{This work was supported by Shanghai Key Laboratory of Scalable Computing and Systems. Lingxiao Jin and Zinuo Cai equally contributed to this work. (Corresponding author: Ruhui Ma).}

}


\maketitle

\begin{abstract}
Serverless computing is increasingly adopted for its ability to manage complex, event-driven workloads without the need for infrastructure provisioning. However, traditional resource allocation in serverless platforms couples CPU and memory, which may not be optimal for all functions. Existing decoupling approaches, while offering some flexibility, are not designed to handle the vast configuration space and complexity of serverless workflows. In this paper, we propose \name, an innovative, automated framework that decouples CPU and memory resources to provide more flexible and efficient provisioning for serverless workloads. \name is composed of two key components: Graph-Centric Scheduler, which identifies critical paths in workflows, and Priority Configurator, which applies priority scheduling techniques to optimize resource allocation. 
Our experimental evaluation demonstrates that \name achieves substantial improvements over state-of-the-art methods, with total search time reductions of 85.8\% and 89.6\%, and cost savings of 49.6\% and 61.7\%, respectively, while maintaining SLO compliance.
\end{abstract}

\begin{IEEEkeywords}
serverless computing, resource configuration, automatization.
\end{IEEEkeywords}

\section{Introduction}

Serverless computing~\cite{li2022serverless, mampage2022holistic} is emerging as a new paradigm in cloud computing, offering distinct advantages over traditional models like Infrastructure-as-a-Service~\cite{manvi2014resource, bhardwaj2010cloud} and Platform-as-a-Service~\cite{al2017elasticity, pahl2015containerization}.
Unlike these earlier services, serverless computing, or Function-as-a-Service (FaaS), breaks down applications into smaller, more manageable functions. This approach significantly reduces the maintenance burden on developers and enables service providers to maximize the use of underlying hardware resources.
Complex applications often require more than one function, which is where serverless workflows come into play~\cite{shahrad2020serverless}.
Visualized as directed acyclic graphs (DAGs), serverless workflows can scale based on demand, providing both operational flexibility and cost-effectiveness. Each function within the workflow operates independently, allowing for efficient resource allocation and management.

Despite its reputation for flexible and granular resource allocation, FaaS is still predicated on user-defined quotas for resource configuration. We characterize existing resource management strategies within FaaS frameworks into three primary categories.
First, memory-centric configurations in AWS Lambda\footnote{\url{https://aws.amazon.com/lambda/pricing/}}, allow developers to set memory quotas, with vCPU and network bandwidth allocated in proportion to the memory allocation.
Second, Google Cloud Functions\footnote{\url{https://cloud.google.com/functions/pricing-1stgen}} providers offer a selection of predefined resource combinations.
Third, there are flexible yet limited configurations, like Alibaba Cloud Function\footnote{\url{https://www.aliyun.com/product/fc/}}, which enforces a memory-to-CPU ratio within a specified range.
It is speculated that these constraints aid cloud providers in the efficient management of resources and potentially reduce the need for extensive physical infrastructure.

However, the existing coupled resource allocation mechanism may not be optimal for all serverless functions.
Bilal \etal~\cite{bilal2023great} highlight the benefits of resource decoupling, which can potentially reduce execution costs by up to 40\% when compared to coupled configurations.
However, their study is limited to individual functions and does not extend to workflows. We analyze three serverless workflows~\cite{mahgoub2022orion} with decoupled CPU-memory resources to evaluate their performance in~\sect{motivation}.
Our findings indicate that a coupled CPU-memory mechanism can result in resource inefficiency and increased costs due to the varying resource requirements of different workflows.
For instance, in the ML Pipeline workflow, a decoupled configuration of 4 vCPUs and 512 MB memory achieves optimal cost by reducing memory usage by 87.5\%, substantially decreasing the overall cost compared to the coupled approach.
%

While decoupling CPU and memory seems a promising strategy for adaptive resource provisioning, it significantly broadens the configuration space, presenting substantial challenges in finding optimal settings.
Previous research~\cite{bilal2023great} suggests a Bayesian optimization-based method for resource allocation, yet this method is geared towards individual functions.
We extend the method to workflows and evaluate it with a Chatbot application in \sect{challenge}.
%
After 100 rounds of sampling, with the algorithm still not converging, we observe a 32.13\% reduction in cost, but the total runtime extends to 9.76 hours, indicating a struggle to reach an optimal solution.
Moreover, the cost exhibits frequent fluctuations and nearly half of these changes are increases.
These challenges have spurred our efforts to develop a more automated, stable, and efficient resource configuration scheme for the decoupling of resources and workflows.

In this paper, we introduce \name, an automated framework designed to configure resources with affinity awareness for serverless workflows. Our key insight lies in the decoupling of computation and storage resources, which enhances flexibility in provisioning resources for serverless workloads.
\name offers three distinct advantages over existing methods.
Firstly, \name surpasses memory-centric allocation schemes~\cite{suresh2020ensure,qiu2022reinforcement,mahgoub2022orion} by decoupling memory and CPU, thereby increasing resource flexibility and efficiency through a comprehensive exploration of serverless workflows' resource affinities.
Secondly, in contrast to peer works utilizing Bayesian optimization for decoupled configurations~\cite{bilal2023great}, our method efficiently manages complex workflows and minimizes the sampling iterations required.
Lastly, \name's resource allocation centers on Service Level Objectives (SLOs), which specify end-to-end latency limits, eliminating manual resource allocation by developers and automating the process.

Specifically, \name integrates Graph-Centric Scheduler and Priority Configurator to optimize serverless workflow costs using a heuristic algorithm.
The Graph-Centric Scheduler starts by breaking down the workflow to find the critical path, which the Priority Configurator then schedules with priority techniques.
This is followed by Graph-Centric Scheduler setting sub-SLOs for related sub-paths and determining resource configurations for functions within them.
Cloud vendors can use \name to find optimal resource configurations that meet end-to-end SLOs upon receiving workflows from developers.
Our experiment illustrates that during searching for the optimal configuration, \name reduces the total runtime by around 85\% compared to SOTA methods. When comparing the optimal configurations, \name achieves cost savings of around 50\% compared to SOTA methods, while still meeting the SLO requirements.

\textbf{Our contributions are highlighted as follows.}

\begin{itemize}
    \item Firstly, we identify that memory-centric resource configuration schemes on mainstream serverless computing platforms are not a one-size-fits-all approach, and existing methods for decoupled resource configuration don't work for workflows because of the larger search space.
    \item Secondly, we propose \name, a framework for decoupled resource configuration in serverless workflows automatically. Leveraging critical path decomposition and priority scheduling, the framework provides a cost-efficient resource allocation strategy with SLO compliance.
    \item Finally, our experiment demonstrates that \name reduces total runtime by 85\% and achieves 50\% cost savings compared to SOTA methods while meeting SLOs. 
\end{itemize}

\begin{figure}[!t]
    \centering
    \includegraphics[width=0.45\textwidth]{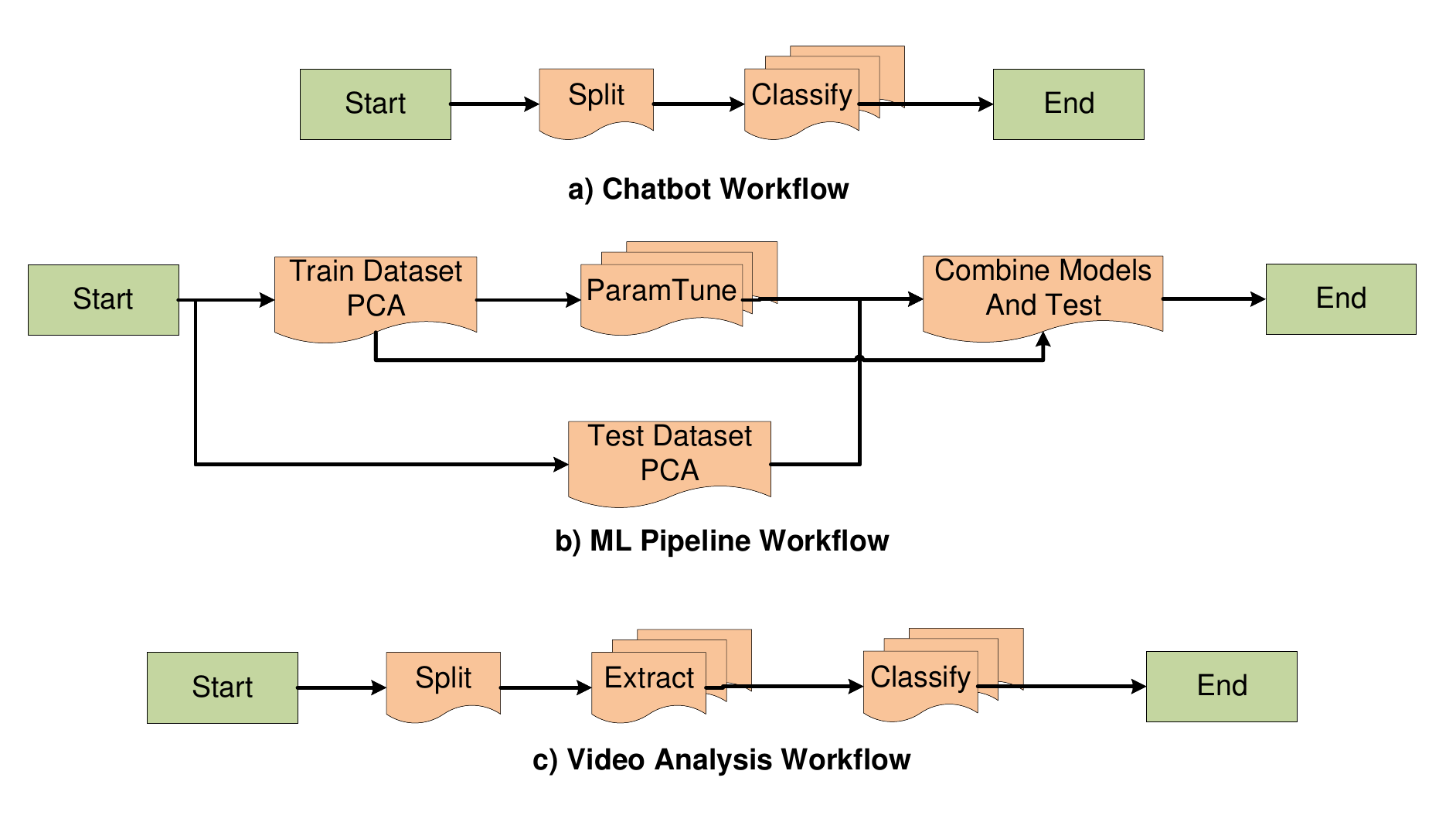}
    \vspace{-0.15in}
    \caption{Architecture of Three Workflows.}
    \label{fig:back_workflows}
    \vspace{-0.2in}
\end{figure}
\begin{figure}[!t]
    \centering
    \begin{subfigure}{0.45\textwidth}
        \centering
        \includegraphics[width=0.95\textwidth]{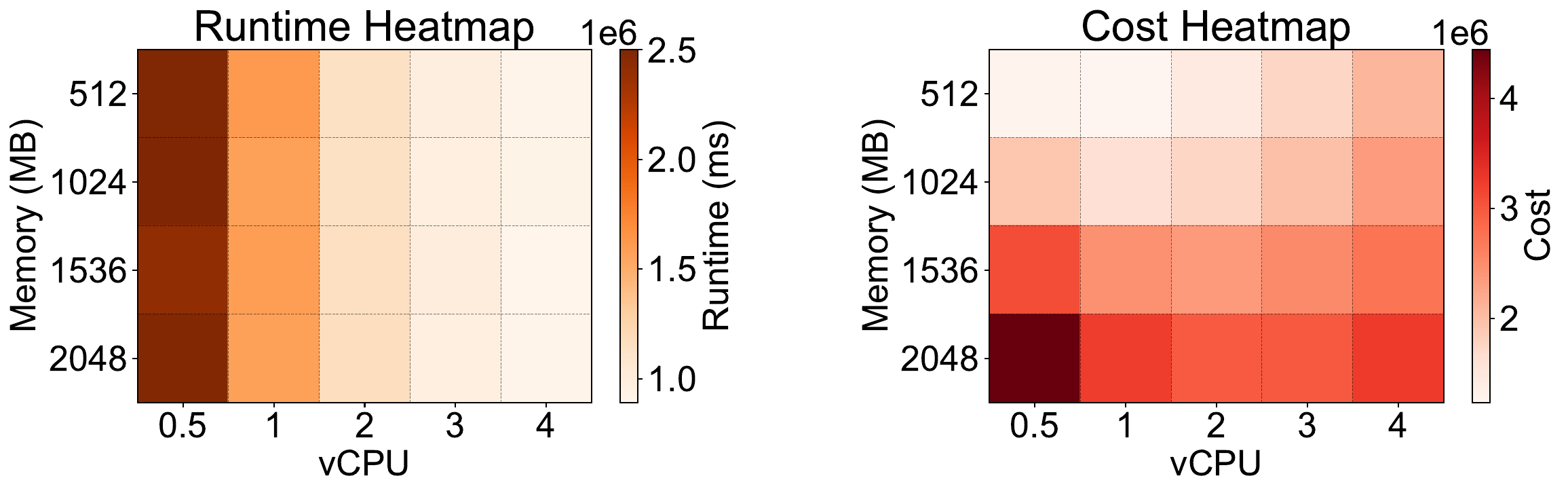}
        \vspace{-0.15in}
        \caption{Chatbot}
        \label{fig:back_chatbot}
    \end{subfigure}
    \begin{subfigure}{0.45\textwidth}
        \centering
        \includegraphics[width=0.95\textwidth]{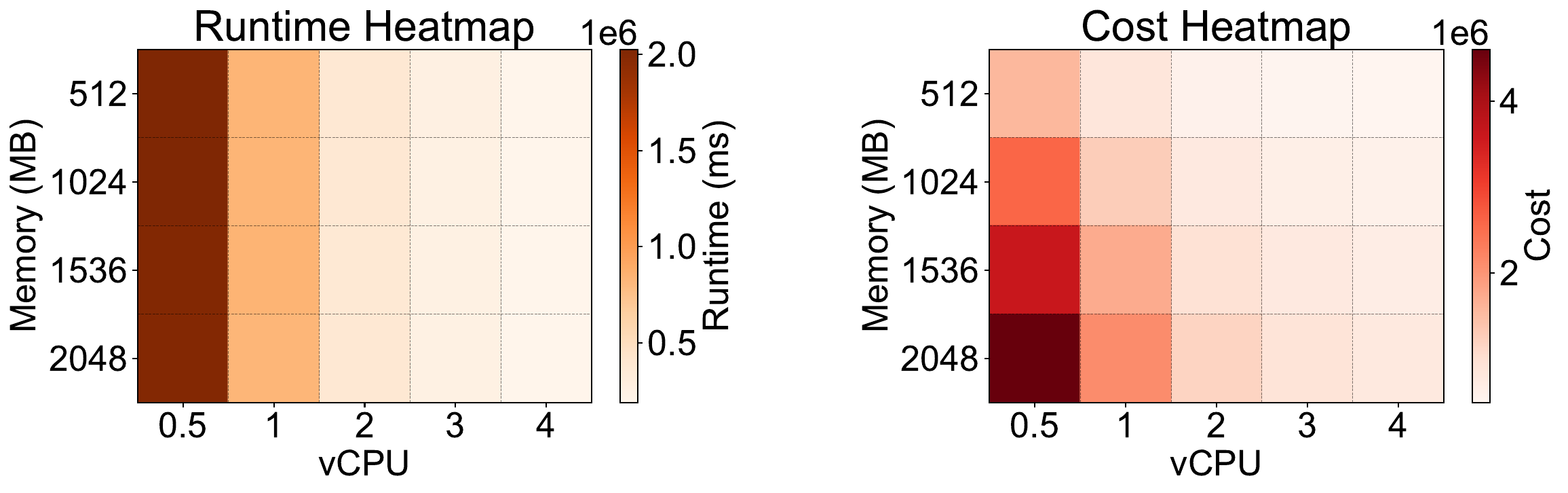}
        \vspace{-0.15in}
        \caption{ML Pipeline}
        \label{fig:back_ml}
        \vspace{0.05in}
    \end{subfigure}
    \begin{subfigure}{0.45\textwidth}
        \centering
        \includegraphics[width=0.95\textwidth]{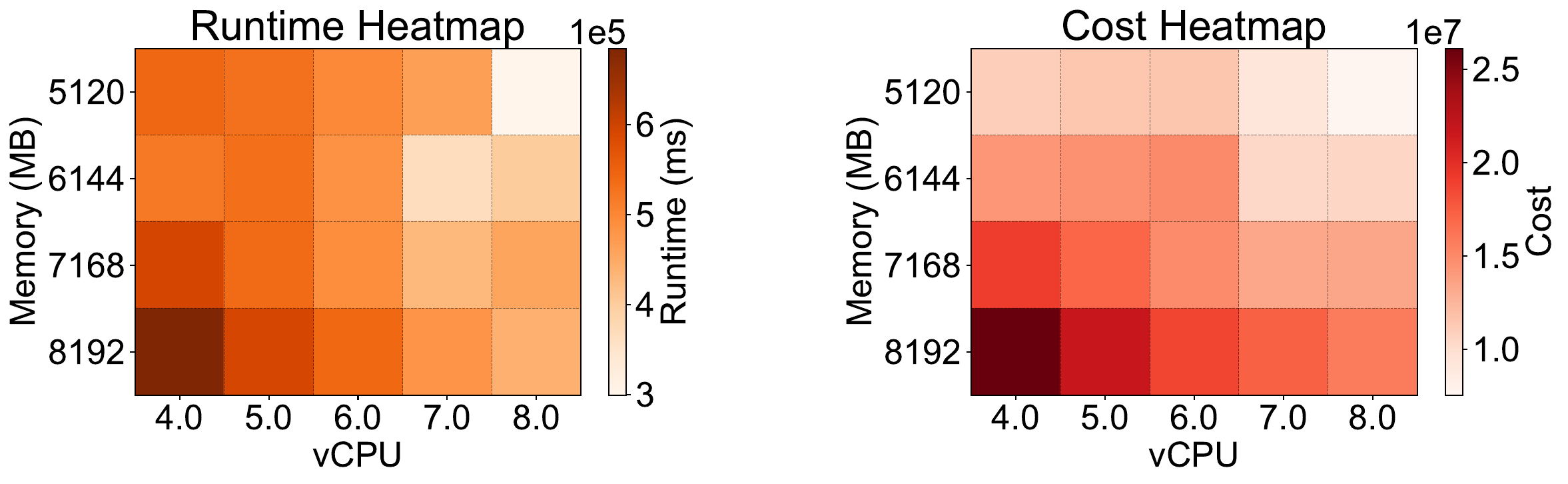}
        \vspace{-0.15in}
        \caption{Video Analysis}
        \label{fig:back_video_heavy}
    \end{subfigure}
    \caption{Runtime and Cost with Decoupled Resources.}
    \label{fig:decoupling}
    \vspace{-0.2in}
\end{figure}
\section{Motivation and Challenge}
\subsection{\textbf{Motivation}: Decoupling for Flexibility}\label{sect:motivation}
While memory-centric resource configurations like AWS Lambda simplify resource management, they may not suit all serverless workloads.
Bilal \etal~\cite{bilal2023great} demonstrate that resource decoupling can reduce execution costs by up to 40\% compared to coupled configurations. 
However, their analysis is restricted to individual functions and does not consider workflows.
To explore the impact of decoupled CPU-memory resources for workflows on runtime and cost, we conduct experiments with three different workflows~\cite{mahgoub2022orion}, Chatbot, ML Pipeline, and Video Analysis. 
\fig{back_workflows} illustrates the architecture of these workflows. The Chatbot workflow processes input, trains classifiers in parallel, and uses remote storage for real-time intent detection. The ML Pipeline workflow achieves machine learning by performing dimensionality reduction, model training, and testing. The Video Analysis workflow splits input videos, extracts key frames, and classifies them.

\fig{decoupling} demonstrates the impact of decoupling CPU and memory on the runtime and cost of the workflows.
\fig{back_chatbot} and \fig{back_ml} show that the runtime for Chatbot and ML Pipeline remains unchanged despite memory variations, indicating inefficiency in memory-centric resource allocation for computation-intensive tasks. Notably, in the ML Pipeline workflow, a decoupled configuration of 4 vCPUs and 512 MB memory achieves the lowest cost, reducing memory usage by 87.5\% compared to the coupled approach, underscoring the need for decoupled strategies.
Additionally, comparing \fig{back_chatbot} and \fig{back_video_heavy} reveals distinct resource affinities across workflows. For example, Chatbot minimizes costs with 512 MB memory and 1 vCPU, while Video Analysis achieves cost efficiency with 5120 MB memory and 8 vCPUs. Thus, resource configurations should be tailored to each workflow to minimize costs while meeting SLOs.
\begin{figure}[!t]
    \centering
    \includegraphics[width=0.4\textwidth]{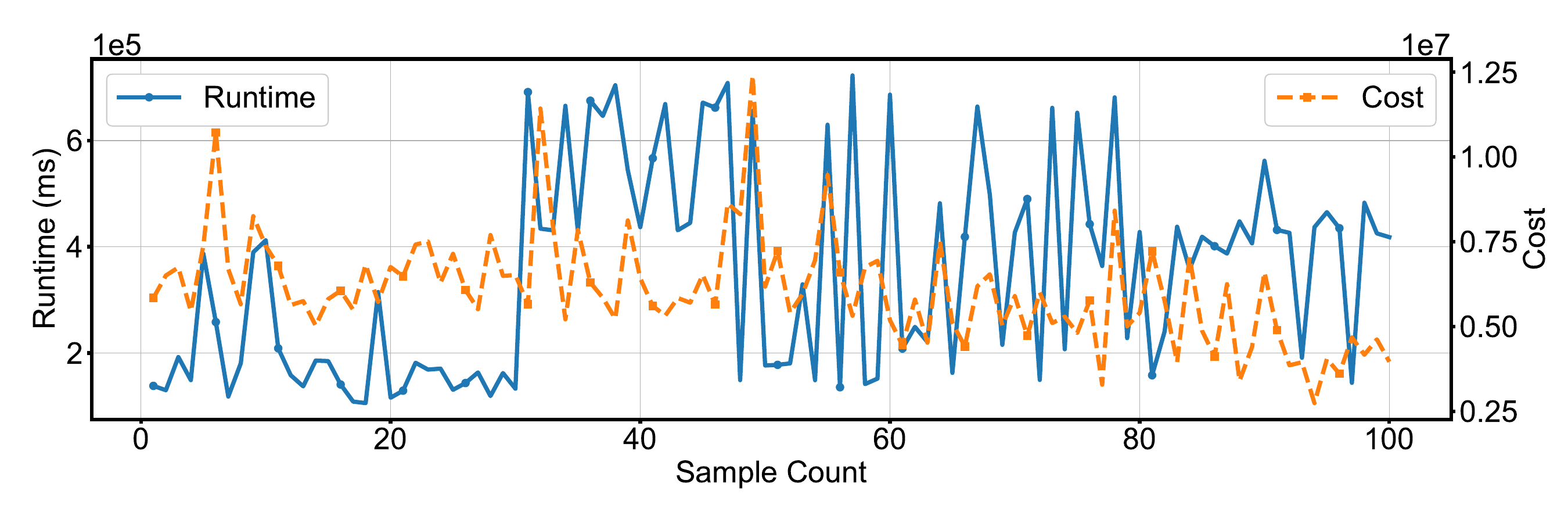}
    \vspace{-0.10in}
    \caption{Bayesian Optimization Search for Chatbot.}
    \label{fig:back_bo}
    \vspace{-0.2in}
\end{figure}

\subsection{\textbf{Challenge}: Larger Configuration Space after Decoupling}\label{sect:challenge}

Decoupling CPU and memory improves resource utilization and reduces costs but significantly expands the configuration space, complicating optimal resource provisioning. Profiling-based methods~\cite{akhtar2020cose, wen2022stepconf, mahgoub2022orion} face increased sampling challenges as configurable resources grow.
The complexity of serverless applications is further exacerbated by the fact that 46\% of applications involve multiple functions~\cite{shahrad2020serverless}.
While prior work~\cite{bilal2023great} explores Bayesian optimization for decoupled resource configuration, it focuses on individual functions, overlooking the unique demands of workflows.
%

We extend the Bayesian Optimization (BO) method from this study to serverless workflows, specifically testing the Chatbot application, to observe changes in runtime and cost with increased sampling.
%
As depicted in \fig{back_bo}, while the cost decreases by 32.13\% over 100 sampling rounds without converging, the total runtime extends to 9.76 hours. 
Additionally, the cost experiences frequent fluctuations, with over half of the changes being increases. 
The average fluctuation amplitude, calculated as the mean absolute difference between consecutive values, accounts for 18.3\% of the mean, which highlights its significant instability.
This instability stems from the expansion of the search space when applying decoupling to workflows, which complicates the identification of the optimal solution using Bayesian Optimization. 
Consequently, current methods struggle to optimize serverless workflows effectively.
\section{Design}

\subsection{Overview}

\begin{figure}[!t]
    \centering
    \includegraphics[width=0.45\textwidth]{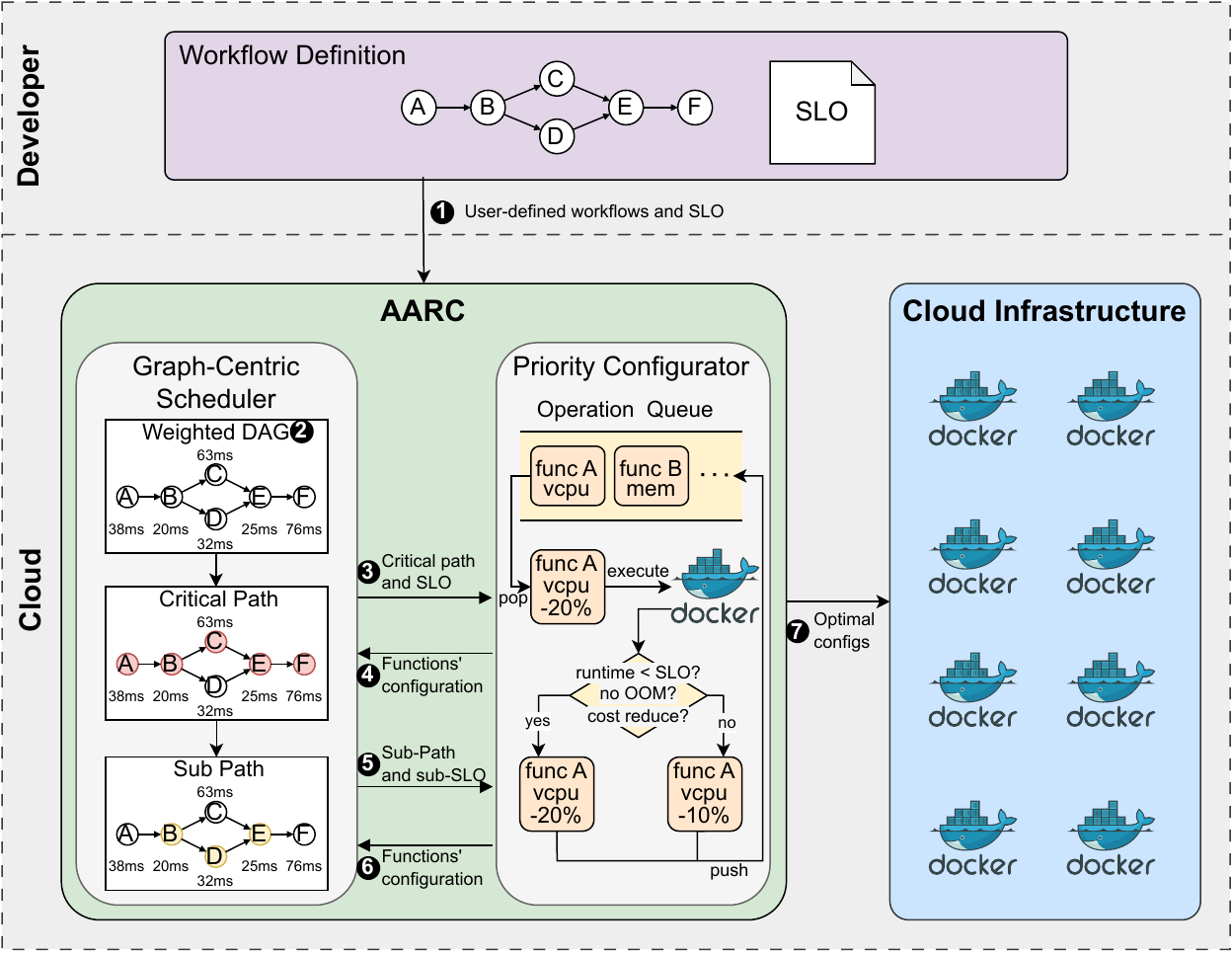}
    \caption{System Architecture. }
    \label{fig:arc}
    \vspace{-0.1in}
\end{figure}

\fig{arc} depicts the overall architecture of \name consisting of two main components: Graph-Centric Scheduler and Priority Configurator.
Graph-Centric Scheduler decomposes the input workflow and identifies its critical path, along with all the sub-paths linked to it.
Priority Configurator configures each function along the critical path and sub-paths.

In our proposed framework, developers submit~\scalebox{1.2}{\ding{182}} their workflow to the cloud platform along with the SLO.
First, the Graph-Centric Scheduler profiles the user-defined workflow based on dummy input, calculating the runtime for each function in the workflow. This runtime is then used as the weight of each node, converting~\scalebox{1.2}{\ding{183}} the workflow into a weighted DAG.
Next, the Scheduler extracts the critical path and its SLO from the weighted DAG and passes~\scalebox{1.2}{\ding{184}} this information to the Priority Configurator.
Priority Configurator then incrementally reduces the memory and CPU allocation for each function through priority-based scheduling to determine~\scalebox{1.2}{\ding{185}} the optimal resource configuration that meets the SLO.
Based on the optimal configuration of the critical path and the SLO, Graph-Centric Scheduler substitutes functions on the non-critical path onto the critical path to generate~\scalebox{1.2}{\ding{186}} sub-paths and corresponding SLOs.
Similarly, Priority Configurator calculates~\scalebox{1.2}{\ding{187}} the optimal resource configuration of the functions in sub-paths, ensuring that the critical path's SLO is not violated.
Finally, Graph-Centric Scheduler finalizes~\scalebox{1.2}{\ding{188}} the resource configuration for subsequent container resource allocation, which ensures SLO compliance with optimal cost efficiency.

\begin{table}[!t]
    \centering
    \caption{Functions in \alg{main} and \alg{sequence}.}
    \label{tab:explain}
    \begin{tabular}{|p{0.15\textwidth}|p{0.25\textwidth}|}
    \hline
    \textbf{Functions} & \textbf{Description} \\ \hline\hline
    \funa{deallocate}{op} & According to the type and step of the given operation $\mathtt{op}$, deprive a portion of the function's resources and return the runtime and cost under this configuration.\\\hline
    \funa{allocate}{op} &  According to type and step of the given operation $\mathtt{op}$, allocate a portion of resources to the function and return a new step and trail of the op. \\\hline
    \funa{find\_critical\_path}{G} & Given a weighted DAG, return the critical path of this DAG. \\\hline 
    \funb{find\_detour\_subpath}{G}{critical\_path} & Given a DAG and its critical path, return all the sub-paths connected to the critical path. \\\hline
    \func{runtime\_sum}{path}{start}{end} & Given the start and the end of a path, calculate the overall duration time between the start and the end. \\\hline
    \end{tabular}
    \vspace{-0.2in}
\end{table}

\subsection{Detailed Design}

\begin{algorithm}[!t]
    \caption{Overall Scheduling}
    \label{alg:main}
    \DontPrintSemicolon
    \SetAlgoLined

    \SetKwComment{Comment}{/* }{ */}
    \SetKwRepeat{Do}{do}{while}
    \SetKwProg{Fn}{Function}{:}{}
    \KwIn{function workflow $\mathtt{G}$, end-to-end latency $\mathtt{SLO}$}
    \KwOut{configurations for each function $\mathtt{G\_configs}$}

    \Fn{\funb{schedule}{G}{SLO}}{
        \Comment{assign base configuration}
        \ForEach{$\mathtt{v_i}\in \mathtt{G}$}{
            $\mathtt{v_i.config} \gets \mathtt{base\_config}$ \;
        }
        \Comment{execute to find critical path}
        execute $\mathtt{G}$ \;
        $\texttt{L} \gets \funa{find\_critical\_path}{G}$ \;
        $\mathtt{G\_configs}\gets\{\}$ \;
        $\mathtt{configs}\gets$ \funb{priority\_configuration}{L}{SLO} \;
        $\mathtt{G\_configs} \gets \mathtt{G\_configs} \cup \mathtt{configs}$ \;
        \Comment{compute configs for subpaths}
        $\mathtt{subpaths} \gets \funb{find\_detour\_subpath}{G}{L}$ \;
        \ForEach{$\mathtt{sp}\in\mathtt{subpaths}$}{
            $\mathtt{SLO^{\prime}} \gets$ \func{runtime\_sum}{L}{sp.start}{sp.end} \;
            \ForEach{$\mathtt{v_i \in sp}$}{
                \If{$\mathtt{v_i}$ \textup{is scheduled}}{
                    $\mathtt{v_i} \gets \funa{pop}{sp}$ \;
                    $\mathtt{SLO^{\prime}} \gets \mathtt{SLO^{\prime}} - \mathtt{v_i.runtime}$ \;
                }
            }
            $\mathtt{configs}\gets$ \funb{priority\_configuration}{sp}{SLO^{\prime}} \;
            $\mathtt{G\_configs} \gets \mathtt{G\_configs} \cup \mathtt{configs}$ \;
        }
        \KwRet{$\mathtt{G\_configs}$}
    }
    \textcode{End Function}
\end{algorithm}

\begin{algorithm}[!t]
    \caption{Priority Configuration}
    \label{alg:sequence}
    \DontPrintSemicolon
    \SetAlgoLined

    \SetKwComment{Comment}{/* }{ */}
    \SetKwRepeat{Do}{do}{while}
    \SetKwProg{Fn}{Function}{:}{}
    \KwIn{function path $\mathtt{L}$, end-to-end latency $\mathtt{SLO}$}
    \KwOut{configuration for each function $\mathtt{configs}$}

    \Fn{\funb{priority\_configuration}{L}{SLO}}{

        $\mathtt{PQ}, \mathtt{count} \gets \fun{priority\_queue}, 0$ \;

        \ForEach{$v_i \in L$}{
            \For{$\mathtt{type} \in \mathtt{[cpu, mem]}$}{
                $\mathtt{func}, \mathtt{priority} \gets v_i, \infty$ \;
                $\mathtt{step}, \mathtt{trial} \gets 1, \mathtt{FUNC\_TRIAL}$ \;
                $\mathtt{op} \gets \{ \mathtt{func}, \mathtt{type}, \mathtt{step}, \mathtt{trail}\}$ \;
                \func{push}{PQ}{op}{priority} \;
            }
        }
        \While{\text{\funa{len}{PQ}} $> 0$ \textcode{and} $\mathtt{count} < \mathtt{MAX\_TRAIL}$}{
            $\mathtt{op}, \mathtt{count} \gets$ \funa{pop}{PQ}, $\mathtt{count}+1$ \;
            $\textup{runtime}, \textup{cost} \gets $\funa{deallocate}{op} \; 
            \eIf{\textup{runtime} $>$ $\mathtt{SLO}$ \textcode{or} \textup{cost increases}}{
                $\mathtt{op.trail}, \mathtt{op.step} \gets $\funa{allocate}{op} \;
                \If{$\mathtt{op.trail} > 0$}{
                    \func{push}{PQ}{op}{0} \;
                }
            }{
                $\mathtt{priority}\gets$ reduced cost \;
                \func{push}{PQ}{op}{priority} \;
            }
        }
        $\mathtt{configs} \gets \{(v_i.cpu, v_i.mem) \mid v_i \in L\}$\;
        \KwRet{$\mathtt{configs}$}
    }
    \textcode{End Function}
\end{algorithm}

Graph-Centric Scheduler employs \textit{Overall Scheduling Algorithm} to process the workflow and SLO, performing critical path-based decomposition of the workflow. It then invokes the Priority Configurator to allocate resources for functions along the paths. Priority Configurator uses \textit{Priority Configuration Algorithm} to handle the function paths and SLO, achieving decoupled resource allocation through priority scheduling.
The pseudocode for the \textit{Overall Scheduling Algorithm} and \textit{Priority Configuration Algorithm} is provided in \alg{main} and \alg{sequence}, with their functions explained in \tab{explain}.

\alg{main} presents the whole scheduling process.
Given a function workflow and the target end-to-end $\mathtt{SLO}$, the algorithm operates as follows: 
Firstly, each function in the workflow is initially assigned an over-provisioned base configuration to ensure the workflow meets the SLO in most cases (Line 2-4).
The workflow is then executed, and the runtime for each function is used as the weight in a DAG to identify the critical path (Line 5-6).
The critical path and the end-to-end SLO are then utilized as inputs for the Priority Configurator, which in turn generates optimized configurations for each function along the critical path (Line 7-9). 
Through a full DAG traversal, the algorithm identifies sub-paths linked to the critical path, defined by their start and end nodes within the critical path, and no intersections with other nodes (Line 10).
For each sub-path, the algorithm calculates the sub-SLO as the time interval between critical path nodes, ensuring critical path consistency during scheduling (Line 12).
The Priority Configurator configures each function along the sub-path, following the same method as with the critical path (Line 19-20).
To prevent conflicts and ensure each function only schedules once, the algorithm sets a scheduled flag for every function. After scheduling a function, the algorithm removes it from the path and adjusts the SLO accordingly (Line 13-18).

\alg{sequence} presents \textit{Priority Configuration Algorithm}, which adopts priority scheduling to optimize the configuration of sequentially executed functions subject to SLO limitation. 
The algorithm initializes a priority queue $\mathtt{PQ}$ to manage operations for each function (Line 2), which requires two distinct operations to adjust CPU and memory quotas (Line 3-10).
Based on the operation type, the algorithm modifies the function's resources, continuously executing operations from $\mathtt{PQ}$ and monitoring their impact on runtime and cost (Line 12-13).
If an operation violates the SLO, increases cost, or encounters an error, the algorithm reverts the resource (Line 14-18). 
Simultaneously, an exponential backoff mechanism reduces the step size to ensure convergence while avoiding over-scheduling (Line 15). 
Operations consistently causing violations ($\mathtt{trail} = 0$) are removed from $\mathtt{PQ}$ (Line 16), while those with potential are re-enqueued with adjusted priorities (Line 17, 20-21).
The loop terminates when $\mathtt{PQ}$ is empty or a user-defined iteration limit, $\mathtt{MAX_{TRAIL}}$, is reached (Line 11).
\section{Performance Evaluation}

\subsection{Experiment Setup}

\paragraph{Environment} 
The experiments run on a machine with 4-socket Intel Xeon Gold 6248R (96 physical core), and 512GB memory. Workflows execute in separate Docker containers, enabling CPU and memory allocation decoupling.

\paragraph{Baselines}
We compare our approach with two baselines: Bayesian Optimization (BO) \cite{bilal2023great} and MAFF gradient descent \cite{zubko2022maff}. Originally designed for single-function resource configuration and memory-centric tasks, these methods have been adapted for workflow optimization.
In the Bayesian Optimization method, configuration parameters are discretized to limit the search space. Memory allocation is available in 64 MB increments from 128 MB to 10,240 MB, while vCPU cores range from 0.1 to 10, independently of memory.
The MAFF gradient descent method iteratively minimizes cost, allocating vCPU cores proportionally (1 core per 1,024 MB of memory). If a workflow's SLO is violated, the process reverts to the previous step and terminates.

\paragraph{Workloads} 
Our experiments utilize three workflows (Chatbot, ML Pipeline, and Video Analysis), with SLOs set to 120s, 120s, and 600s, respectively.
These applications capture key characteristics of serverless DAGs. Specifically, Video Analysis and Chatbot exhibit a scatter communication pattern, while ML Pipeline follows a broadcast pattern \cite{mahgoub2022orion}. 
%

\paragraph{Metrics}
We evaluate performance using runtime and cost metrics, extending the pricing model of AWS Lambda to decoupled resources.
Let $\mathtt{cost_{ij}}$ represent the cost of serverless function $\mathtt{v_i}$ configured to $\mathtt{(cpu_j, mem_j)}$ with runtime $\mathtt{t_{ij}}$. We denote the price per vCPU-second as $\mathtt{\mu_0}$, the price per GB-second as $\mathtt{\mu_1}$, and the price for function requests and orchestration as $\mathtt{\mu_2}$. All are constants. The cost equation is then $\mathtt{cost_{ij} = t_{ij}(\mu_0 \cdot cpu_j + \mu_1 \cdot mem_j) + \mu_2}$.  Here, $\mu_0$, $\mu_1$, and $\mu_2$ are set to 0.512, 0.001, and 0, respectively.

\subsection{Effectiveness of Configuration Search}
In the configuration search experiment, we set the initial configurations for each workflow. 
%
Then, we perform the configuration search using three methods: \name, Bayesian Optimization, and MAFF.

\paragraph{Overall Efficiency}

\begin{figure}[t]
    \centering
    \begin{subfigure}{0.24\textwidth}
        \centering
        \includegraphics[width=0.95\textwidth]{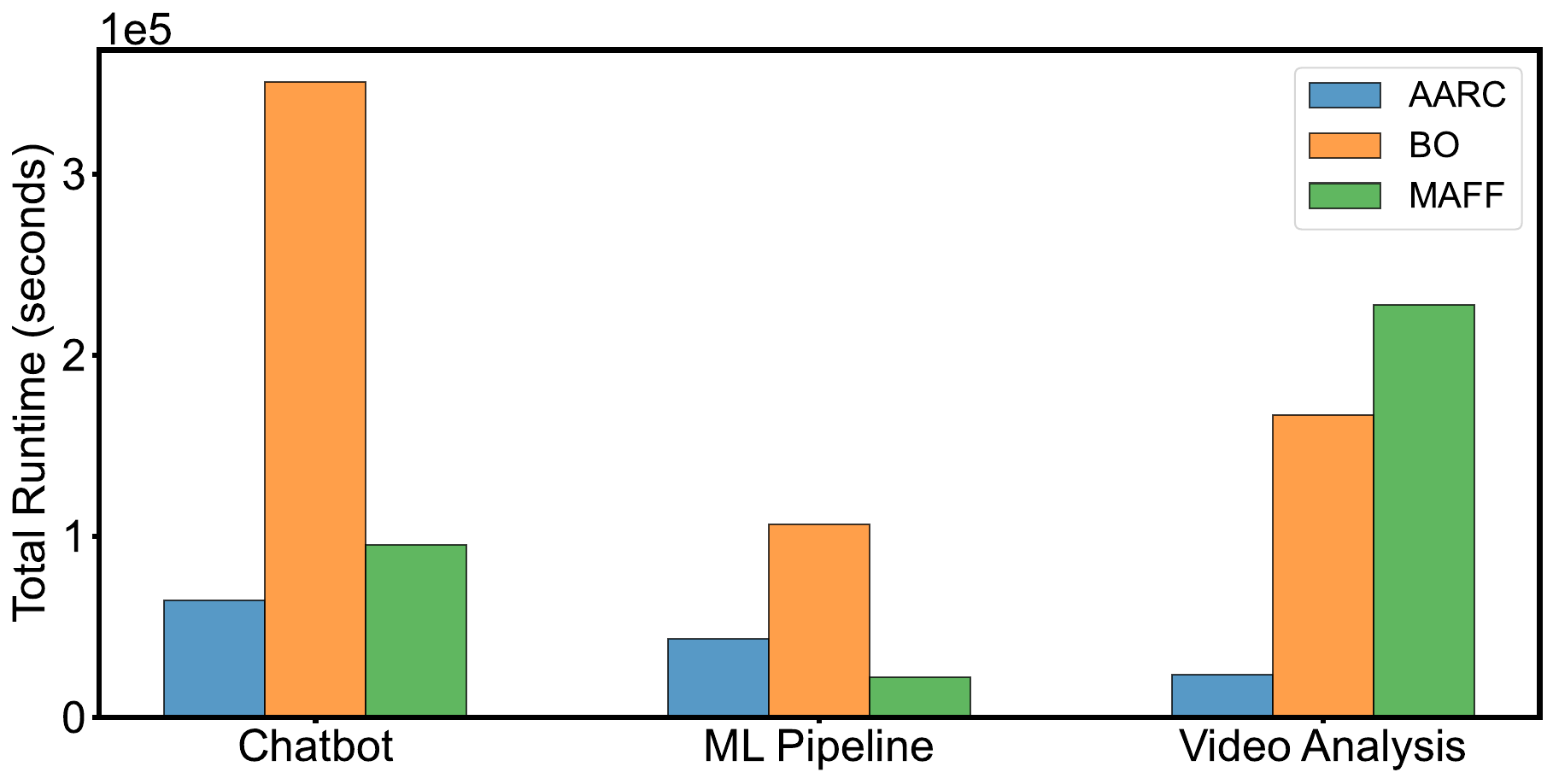}
        \vspace{-0.05in}
        \caption{Total Runtime}
        \label{fig:eva_overall_runtime}
    \end{subfigure}
    \begin{subfigure}{0.24\textwidth}
        \centering
        \includegraphics[width=0.95\textwidth]{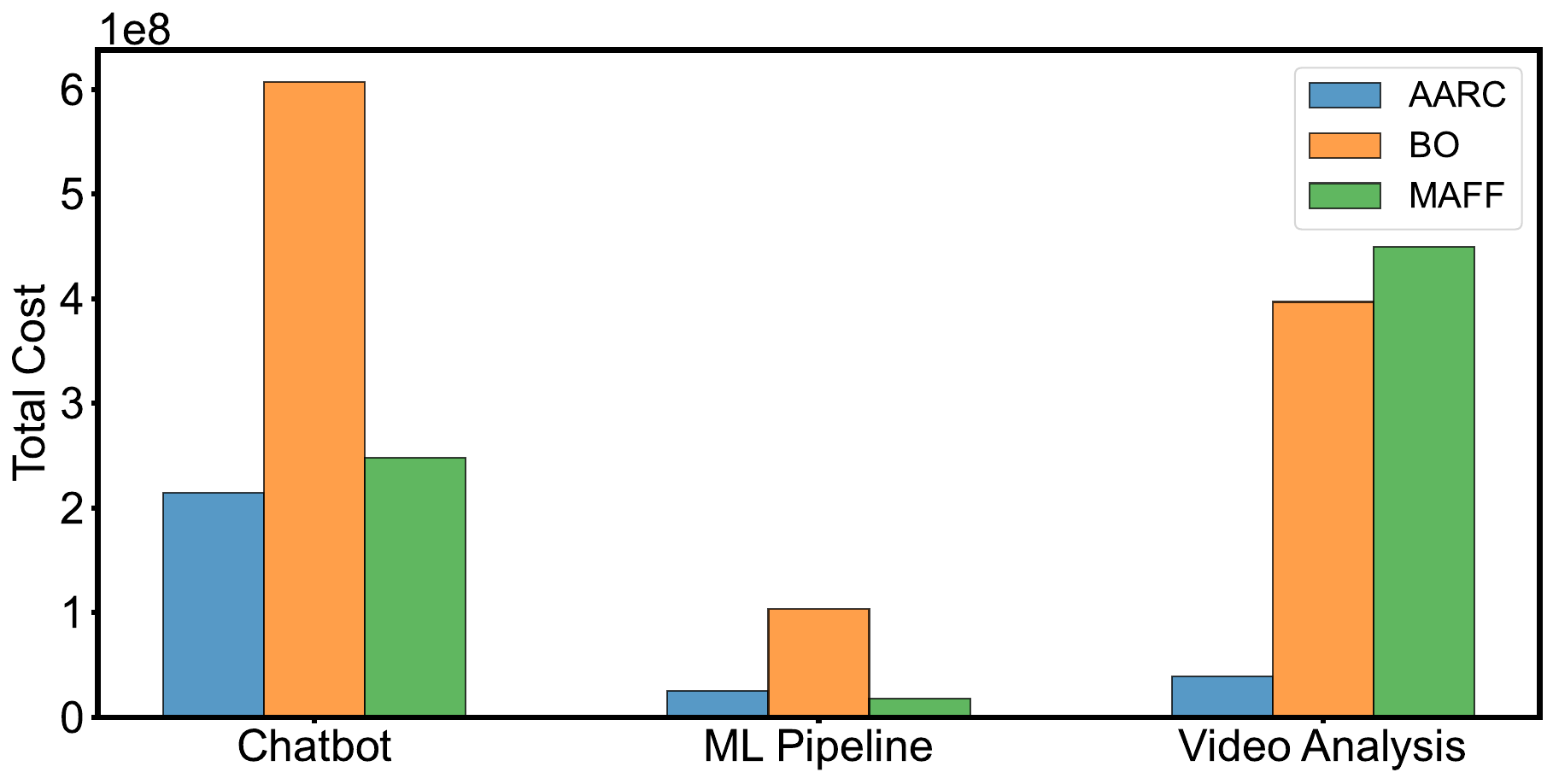}
        \vspace{-0.05in}
        \caption{Total Cost}
        \label{fig:eva_overall_cost}
    \end{subfigure}
    \caption{Overall Sample Cost and Runtime Comparison.}
    \label{fig:eva_search}
    \vspace{-0.1in}
\end{figure}

\begin{table}[!t]
    \centering
    \captionof{table}{Average Runtime and Cost Comparison.}
    \label{tab:eva_simulate_cost}
    \resizebox*{0.48\textwidth}{!}{
        \begin{tabular}{|c|c|c|c|c|c|c|}
        \hline
        & \multicolumn{2}{c|}{\textbf{Chatbot}} & \multicolumn{2}{c|}{\textbf{ML Pipeline}} & \multicolumn{2}{c|}{\textbf{Video Analysis}} \\ \hline
        & Runtime (s) & Cost & Runtime (s) & Cost & Runtime (s) & Cost\\ \hline\hline
        \textbf{AARC} & 103.7±3.2 & 2390.9k & 77.1±2.6  & 435.0k  & 316.8±6.6  & 53.6k \\ \hline
\textbf{BO}   & 114.7±1.9 & 4275.2k & 60.0±0.7  & 863.5k  & 519.9±8.3  & 82.4k \\ \hline
\textbf{MAFF} & 115.3±3.1 & 3477.5k & 109.5±2.0 & 1136.6k & 578.2±19.3 & 98.8k \\ \hline
        \end{tabular}
    }
    \vspace{-0.2in}
\end{table}

\fig{eva_search} shows the total runtime and cost of the sampling process.
In all workflows, \name outperforms the Bayesian Optimization, especially in the Video Analysis workflow, reducing runtime by 85.8\% and cost by 90.1\%. 
This improvement stems from \name's priority scheduling strategy for decoupled resources, which expands the configuration search space while requiring fewer samples.
In the Chatbot workflow, \name performs 64 samples, slightly more than MAFF's 61, but achieves a 31.9\% reduction in runtime and 13.4\% in cost. 
%
This is because MAFF's proportional allocation scheme reduces the search space but risks local optima, leading to higher runtime and costs due to resource coupling.
In the Video Analysis workflow, \name reduces runtime and cost by 89.6\% and 91.3\% compared to MAFF. 
However, in the ML Pipeline workflow, \name samples 50 times compared to MAFF's 15, resulting in lower runtime and cost for MAFF. 
This is due to the ML Pipeline's high CPU and low memory demands, where proportional adjustments often hit local optima, requiring fewer samples to meet exit conditions.

\begin{figure*}
    \centering
    \begin{subfigure}{0.32\textwidth}
        \centering
        \includegraphics[width=0.95\textwidth]{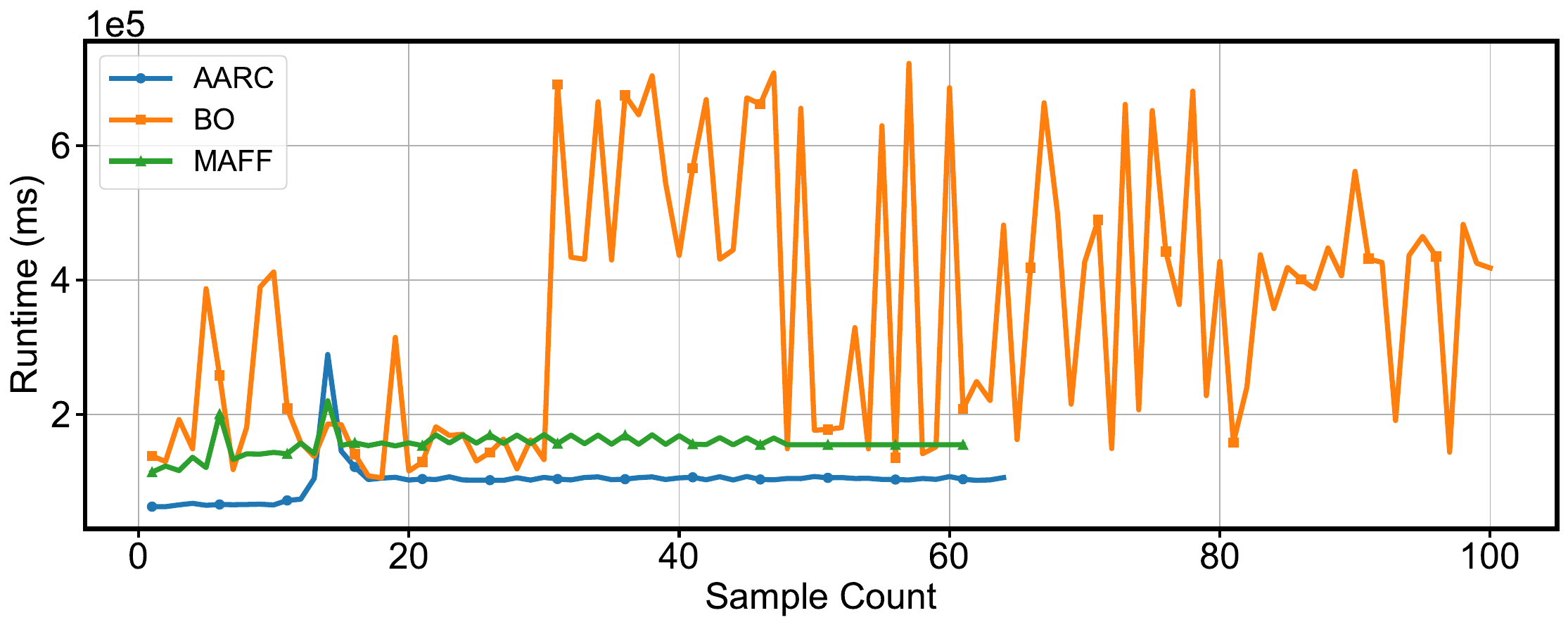}
        \vspace{-0.05in}
        \caption{Chatbot}
        \label{fig:eva_runtime_efficiency_chatbot}
    \end{subfigure}
    \hfill
    \begin{subfigure}{0.32\textwidth}
        \centering
        \includegraphics[width=0.95\textwidth]{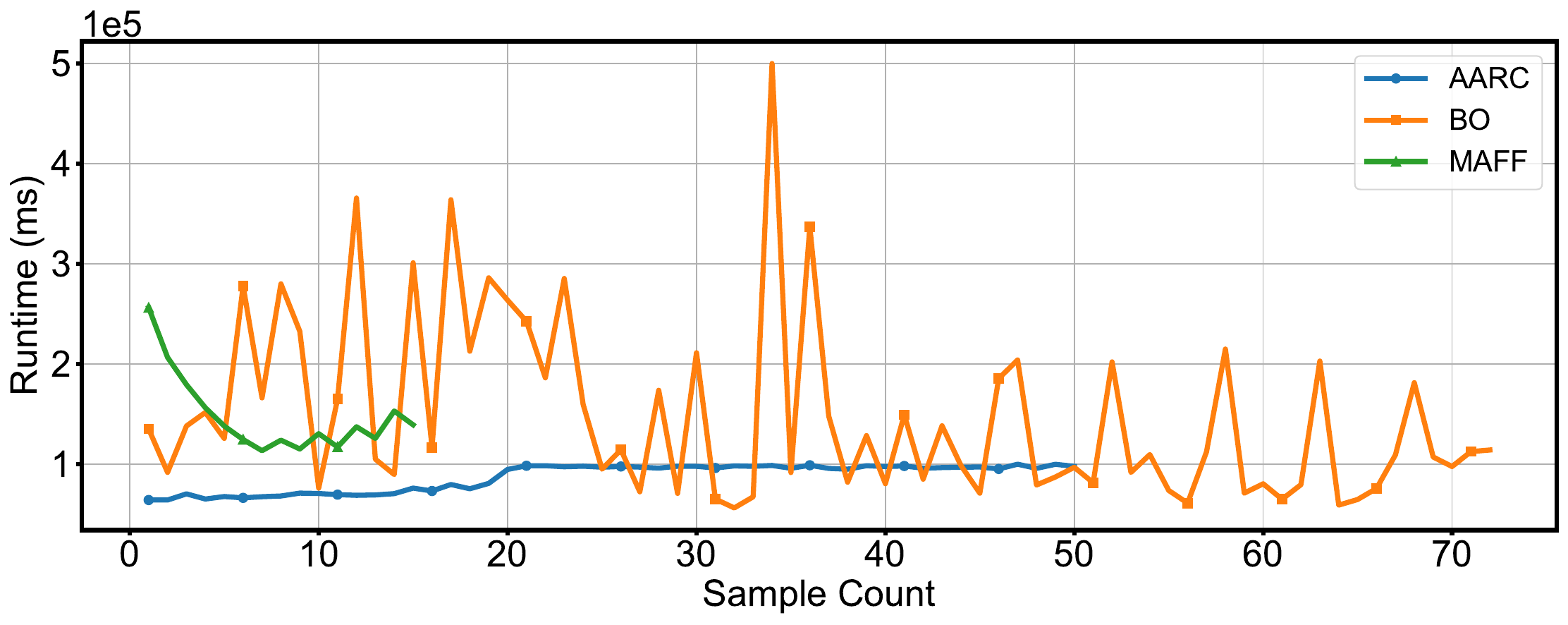}
        \vspace{-0.05in}
        \caption{ML Pipeline}
        \label{fig:eva_runtime_efficiency_mlPipeline}
    \end{subfigure}
    \hfill
    \begin{subfigure}{0.32\textwidth}
        \centering
        \includegraphics[width=0.95\textwidth]{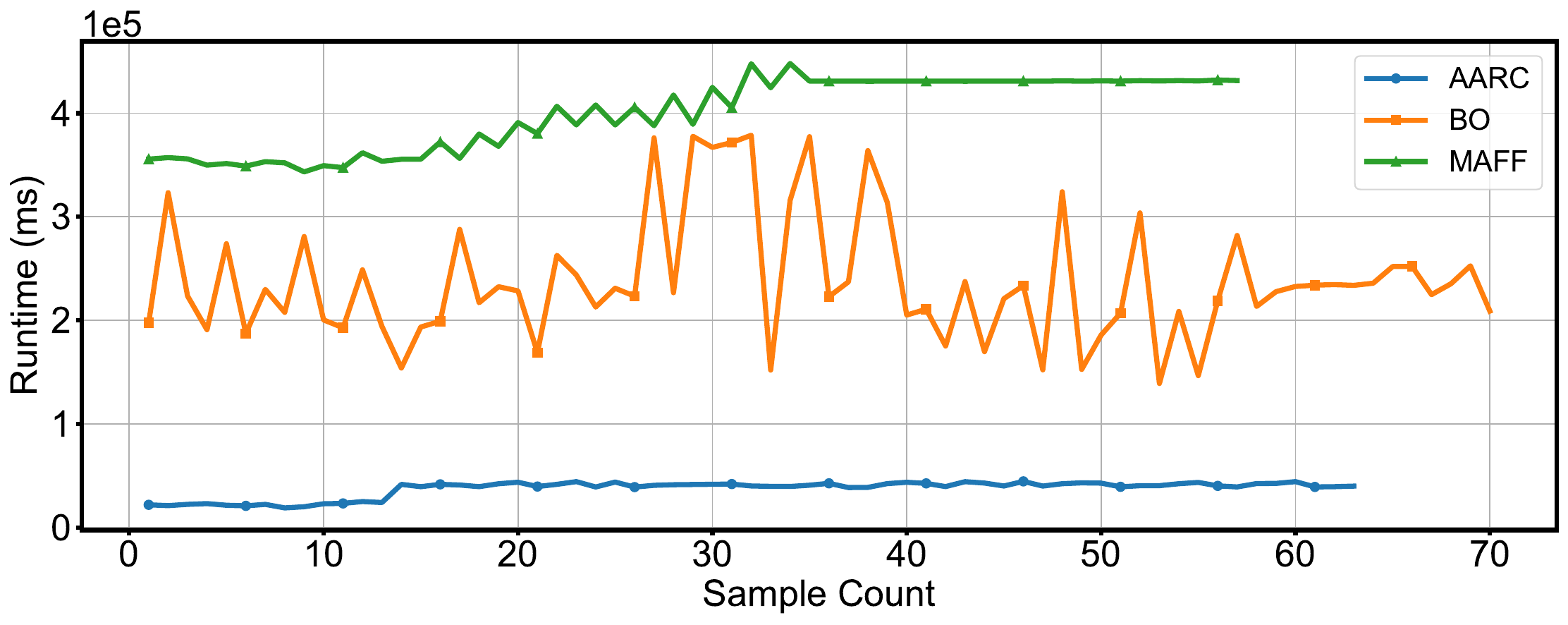}
        \vspace{-0.05in}
        \caption{Video Analysis}
        \label{fig:eva_runtime_efficiency_video}
    \end{subfigure}
    \vspace{-0.05in}
    \caption{Runtime Changing with Sample Counts of Different Methods under Different Workflows.}
    \label{fig:eva_efficiency_runtime}
    \vspace{-0.15in}
\end{figure*}

\begin{figure*}
    \centering
    \begin{subfigure}{0.32\textwidth}
        \centering
        \includegraphics[width=0.95\textwidth]{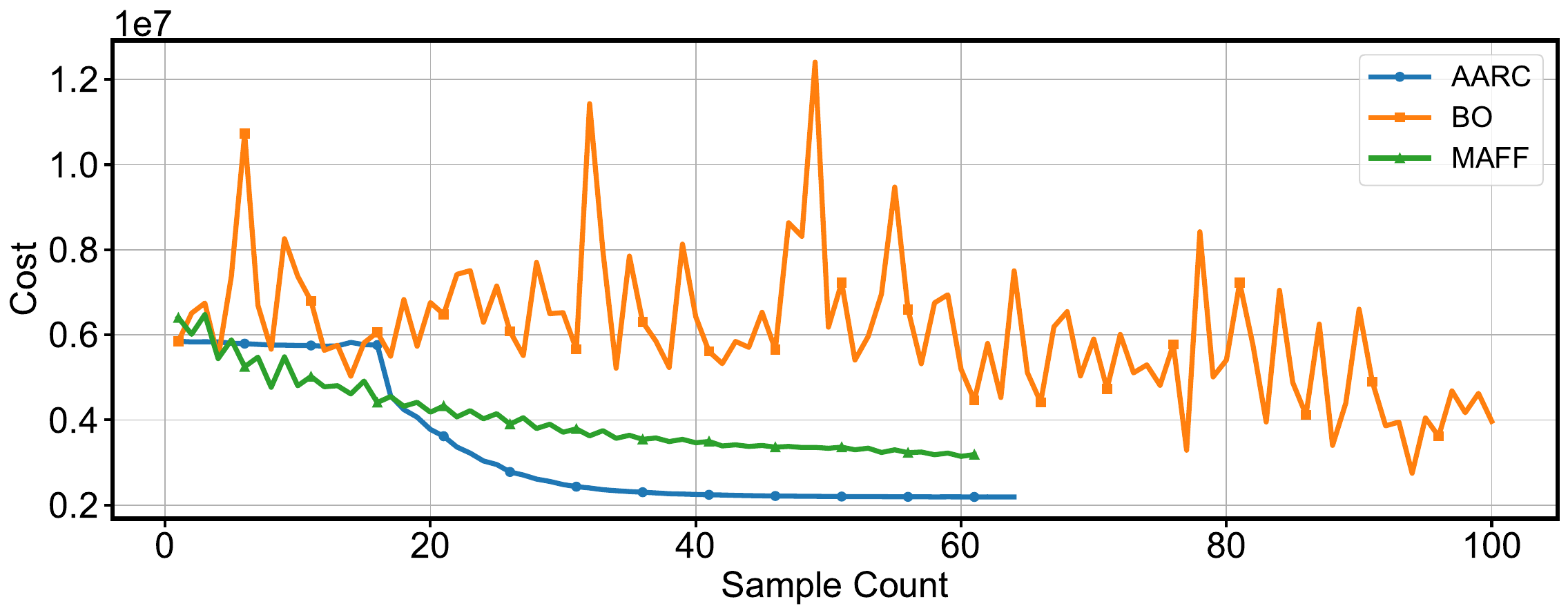}
        \vspace{-0.05in}
        \caption{Chatbot}
        \label{fig:eva_cost_efficiency_chatbot}
    \end{subfigure}
    \hfill
    \begin{subfigure}{0.32\textwidth}
        \centering
        \includegraphics[width=0.95\textwidth]{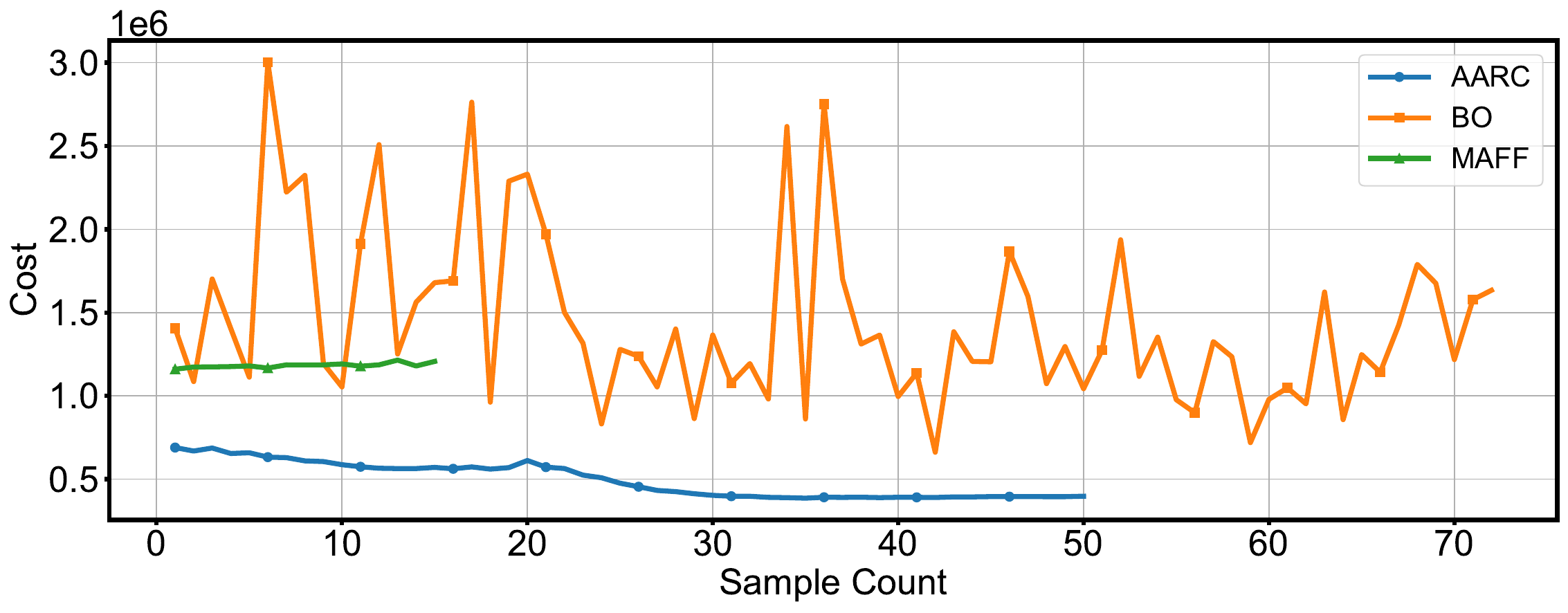}
        \vspace{-0.05in}
        \caption{ML Pipeline}
        \label{fig:eva_cost_efficiency_mlPipeline}
    \end{subfigure}
    \hfill
    \begin{subfigure}{0.32\textwidth}
        \centering
        \includegraphics[width=0.95\textwidth]{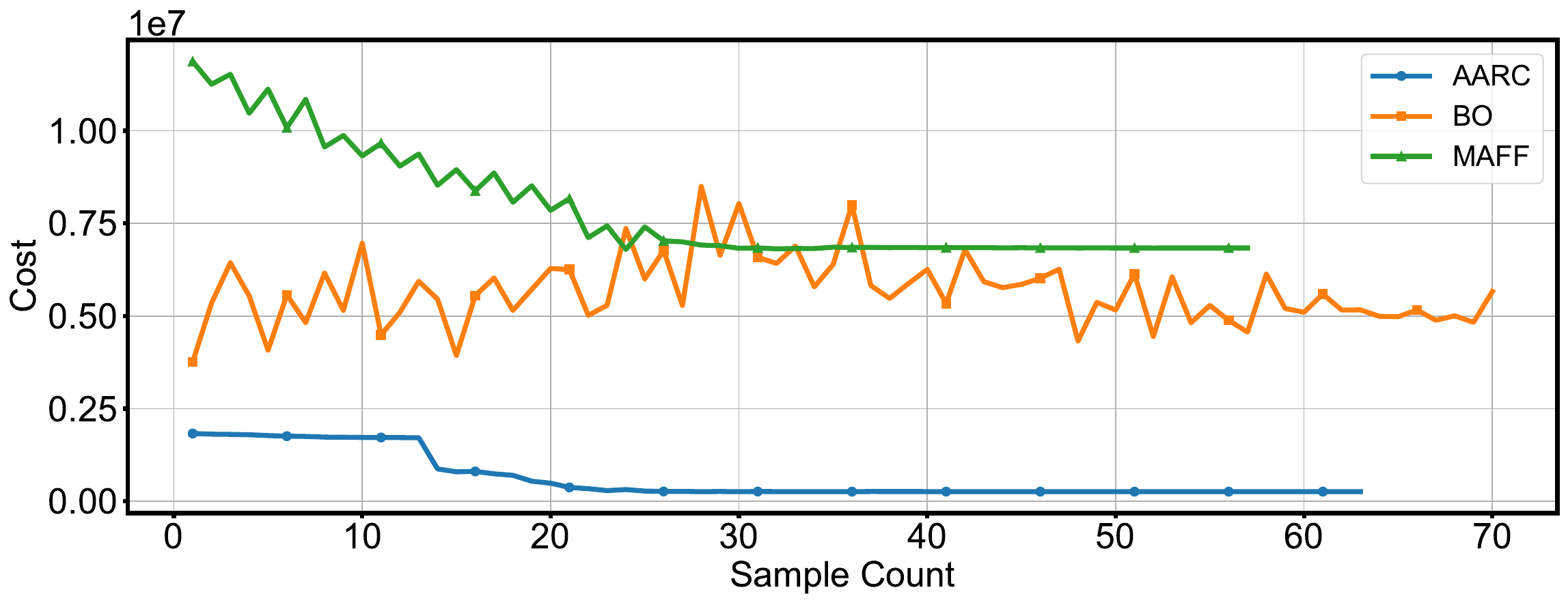}
        \vspace{-0.05in}
        \caption{Video Analysis}
        \label{fig:eva_cost_efficiency_video}
    \end{subfigure}
    \vspace{-0.05in}
    \caption{Cost Changing with Sample Counts of Different Methods under Different Workflows.}
    \label{fig:eva_efficiency_cost}
    \vspace{-0.2in}
\end{figure*}

\paragraph{Sampling Efficiency}
We further illustrate the effectiveness of optimal configurations calculated by each method as the sampling count increases, evaluating these outcomes through workflow runtime and cost.
\fig{eva_efficiency_runtime} shows the change in runtime with sampling count under different configurations for each workflow. 
Our goal is to minimize cost while meeting the SLO, so runtime shows an upward trend using \name. 
The Bayesian Optimization method has the highest sampling count and significant instability, as its efficiency decreases due to the large search space created by resource decoupling. 
In contrast, our priority scheduling strategy maintains search efficiency while accelerating convergence.
\fig{eva_efficiency_cost} depicts the change in cost with sampling count under different configurations. 
Using \name, cost shows a downward trend and converges with fewer samples. 
In the ML Pipeline workflow, the MAFF method quickly falls into local optima due to its coupled resource configuration search approach, making it difficult to discover more cost-effective configurations. 

\subsection{Performance of Optimal Configuration}
To validate the performance of \name in meeting the SLO requirements and execution cost, we execute the workflow 100 times using the configurations generated by the aforementioned methods and calculate its average runtime and cost. 





\begin{figure}
    \centering
    \setcounter{subfigure}{0}
    \begin{subfigure}{0.24\textwidth}
        \includegraphics[width=0.98\textwidth]{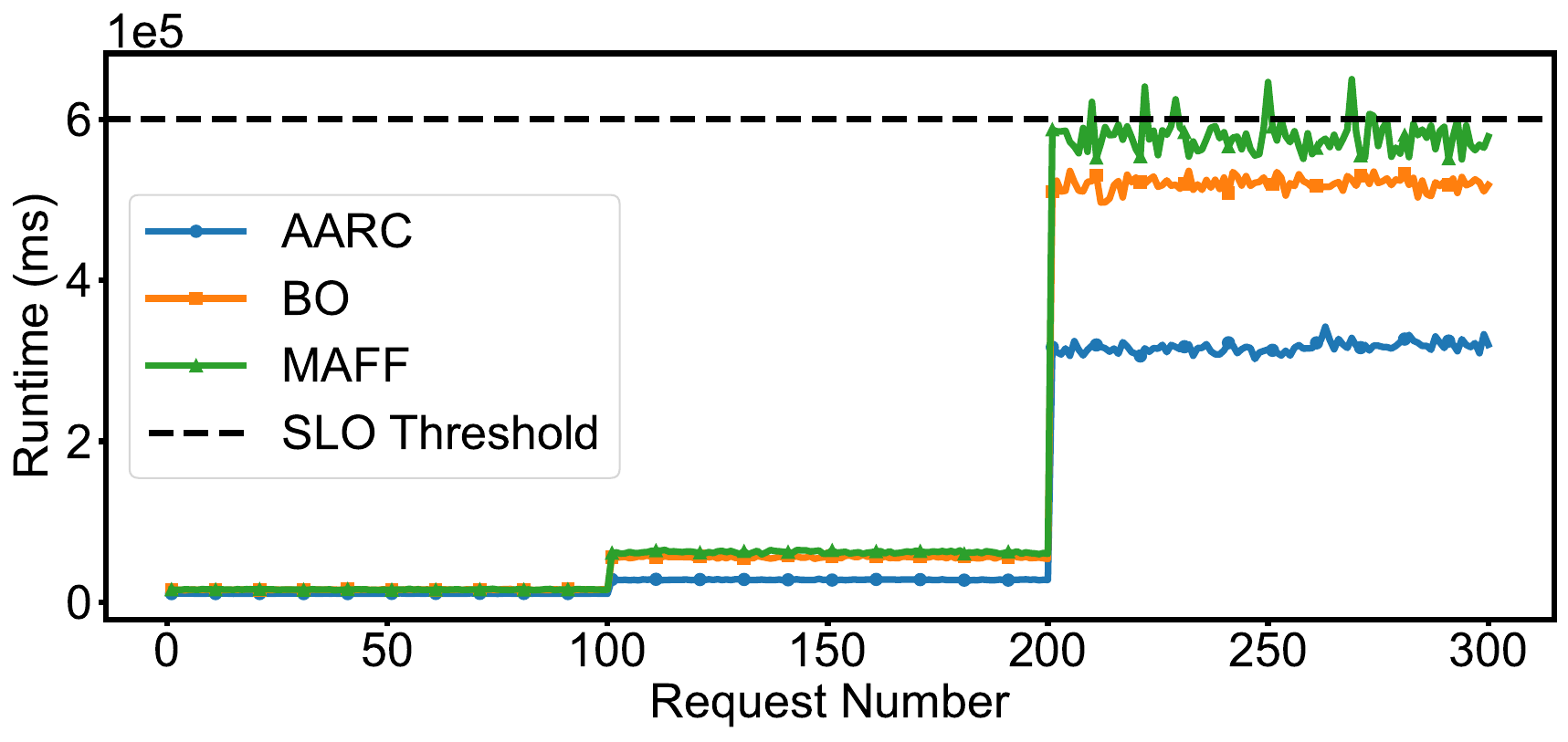}
        \vspace{-0.05in}
        \caption{Runtime}
        \label{fig:diss_video_slo}
    \end{subfigure}
    \hfill
    \begin{subfigure}{0.24\textwidth}
        \includegraphics[width=0.98\textwidth]{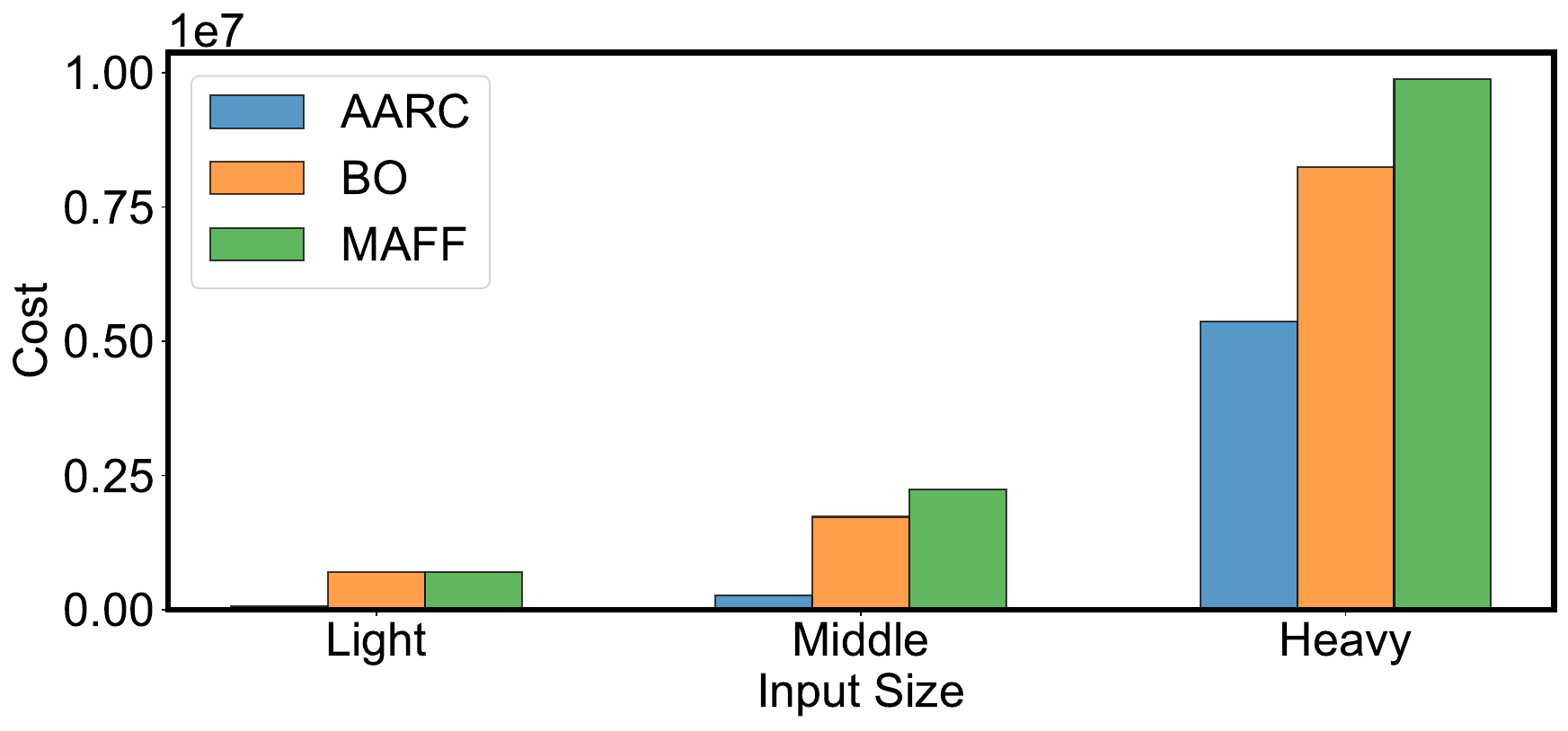}
        \vspace{-0.05in}
        \caption{Cost}
        \label{fig:diss_video_cost}
    \end{subfigure}
    \caption{Performance Across Input Sizes in Video Analysis.}
    \vspace{-0.2in}
\end{figure}

\paragraph{SLO Violation} 
All methods meet the SLO constraints, with the average runtime and standard deviation shown in \tab{eva_simulate_cost}. 
\name satisfies SLO because the algorithm reverts resources during configuration search when SLO violations occur and incorporates sub-paths into the critical path, ensuring both critical path consistency and SLO compliance. 
This demonstrates the effectiveness of our approach in meeting SLO requirements and its potential for integration into real-world systems with high reliability and performance standards.

\paragraph{Cost Reduction} \tab{eva_simulate_cost} shows the execution cost of workflows with configurations found by different methods. 
In the Chatbot, ML Pipeline, and Video Analysis workflows, \name reduces costs significantly compared to Bayesian Optimization and MAFF. 
The reductions are 44.0\% and 31.2\% for Chatbot, 49.6\% and 61.7\% for ML Pipeline, and 34.9\% and 45.7\% for Video Analysis, respectively.
Compared to the Bayesian Optimization method, our approach uses priority scheduling, which allows for more stable identification of suitable decoupled resource configurations. 
Unlike MAFF's proportional configuration method, our approach achieves decoupled resource allocation, resulting in lower-cost configurations. 
Since the ML Pipeline has high CPU demands and low memory demands, decoupling resources yields better results.

\subsection{\textbf{Discussion}: Input-Aware Configuration}
%
Considering that workflow execution efficiency can be input-sensitive, we further enhanced our design by adding an Input-Aware Configuration Engine Plugin. 
The Video Analysis workflow is input-sensitive, where different input video sizes correspond to different optimal configurations.
%
%
If developers trigger the plugin, the Engine analyzes the characteristics of the input data, such as video bitrate and duration. 
Based on the identified features, the Engine sorts the inputs and invokes Graph-Centric Scheduler and Priority Configurator to determine the optimal resource configuration scheme for each input. 
When a request arrives, the Engine analyzes the input scale and allocates the input to different configurations. 

To validate our design, we evaulate Video Analysis workflow using three input sizes: light, middle, and heavy. 
\fig{diss_video_slo} shows workflow runtime with light, middle, and heavy inputs in sequence. 
Since MAFF does not adjust configurations for input-sensitive workflows, it may violate the SLO under heavy inputs, while our method consistently remains within SLO limits.
\fig{diss_video_cost} shows the average cost of the workflow under the three input sizes. Because our method can select configurations based on input size, it optimizes cost by 89.9\% and 89.8\% compared to MAFF and Bayesian Optimization under light input, and by 45.7\% and 34.9\% under heavy input.


\section{Related Work}

\paragraph{Serverless Resource Configuration}
Configuring resources in serverless computing involves optimizing memory and CPU allocations to balance performance and cost. Offline approaches like MAFF~\cite{zubko2022maff} and COSE~\cite{akhtar2020cose} rely on optimization algorithms and performance modeling to determine configurations. 
%
%
Bilal \etal~\cite{bilal2023great} explores decoupling memory and CPU configurations, leveraging diverse VM types to enhance optimization but still faces challenges with workload variability.
To address such dynamism, online configuration methods have emerged. Sizeless~\cite{eismann2021sizeless} uses real-time performance data and prediction models to adjust resources dynamically. FaaSDeliver~\cite{yu2023faasdeliver} further refines this by monitoring per-invocation metrics and employing advanced estimators for resource allocation. 
However, these methods introduce overhead from continuous monitoring and data processing, potentially inflating costs for serverless applications.

\paragraph{Serverless Workflows Optimization}
Due to the prevalence of workflows in serverless workloads, workflow configuration optimization has gained attention from academia and industry~\cite{li2022faasflow, li2023dataflower, wang2024probabilistic}.
SLAM~\cite{safaryan2022slam} uses distributed tracing to detect relationships between functions and leverages execution times under different memory configurations to estimate overall application execution time.
Raza \etal~\cite{raza2023configuration} improve COSE for serverless workflows, optimizing performance and cost across the entire workflow.
StepConf~\cite{wen2022stepconf} provides dynamic memory allocation for serverless workflows, considering intra-function and inter-function parallelism to fully utilize the parallel capacity of serverless functions.
%
%
%
However, these methods still do not effectively decouple resources, leaving room for further optimization.
\section{Conclusion}

We propose \name, an affinity-aware resource configuration framework for serverless workflows, capitalizing on the concept of resource decoupling within the serverless paradigm. 
By leveraging our framework, developers are relieved of the burden of manual serverless workflow configuration. 
Experimental results indicate that, during the process of searching for the optimal configuration, \name achieves a total runtime reduction of 85.8\% and 89.6\% compared to Bayesian Optimization and MAFF, respectively. Additionally, it reduces resource usage by 49.6\% and 61.7\% while satisfying the SLO requirements.

\bibliographystyle{IEEEtran}
\bibliography{main}

\begin{thebibliography}{10}
\providecommand{\url}[1]{#1}
\csname url@samestyle\endcsname
\providecommand{\newblock}{\relax}
\providecommand{\bibinfo}[2]{#2}
\providecommand{\BIBentrySTDinterwordspacing}{\spaceskip=0pt\relax}
\providecommand{\BIBentryALTinterwordstretchfactor}{4}
\providecommand{\BIBentryALTinterwordspacing}{\spaceskip=\fontdimen2\font plus
\BIBentryALTinterwordstretchfactor\fontdimen3\font minus \fontdimen4\font\relax}
\providecommand{\BIBforeignlanguage}[2]{{%
\expandafter\ifx\csname l@#1\endcsname\relax
\typeout{** WARNING: IEEEtran.bst: No hyphenation pattern has been}%
\typeout{** loaded for the language `#1'. Using the pattern for}%
\typeout{** the default language instead.}%
\else
\language=\csname l@#1\endcsname
\fi
#2}}
\providecommand{\BIBdecl}{\relax}
\BIBdecl

\bibitem{li2022serverless}
Z.~Li, L.~Guo, J.~Cheng, Q.~Chen, B.~He, and M.~Guo, ``The serverless computing survey: A technical primer for design architecture,'' \emph{ACM Computing Surveys}, vol.~54, no. 10s, pp. 1--34, 2022.

\bibitem{mampage2022holistic}
A.~Mampage, S.~Karunasekera, and R.~Buyya, ``A holistic view on resource management in serverless computing environments: Taxonomy and future directions,'' \emph{ACM Computing Surveys}, vol.~54, no. 11s, pp. 1--36, 2022.

\bibitem{manvi2014resource}
S.~S. Manvi and G.~K. Shyam, ``Resource management for infrastructure as a service ({IaaS}) in cloud computing: A survey,'' \emph{Journal of Network and Computer Applications}, vol.~41, pp. 424--440, 2014.

\bibitem{bhardwaj2010cloud}
S.~Bhardwaj, L.~Jain, and S.~Jain, ``Cloud computing: A study of infrastructure as a service (iaas),'' \emph{International Journal of engineering and information Technology}, vol.~2, no.~1, pp. 60--63, 2010.

\bibitem{al2017elasticity}
Y.~Al-Dhuraibi, F.~Paraiso, N.~Djarallah, and P.~Merle, ``Elasticity in cloud computing: state of art and research challenges,'' \emph{IEEE Transactions on Service Computing}, vol.~11, no.~2, pp. 430--447, 2017.

\bibitem{pahl2015containerization}
C.~Pahl, ``Containerization and the paas cloud,'' \emph{IEEE Cloud Computing}, vol.~2, no.~3, pp. 24--31, 2015.

\bibitem{shahrad2020serverless}
M.~Shahrad, R.~Fonseca, {\'I}.~Goiri, G.~Chaudhry, P.~Batum, J.~Cooke, E.~Laureano, C.~Tresness, M.~Russinovich, and R.~Bianchini, ``Serverless in the wild: Characterizing and optimizing the serverless workload at a large cloud provider,'' in \emph{Proceedings of USENIX Annul Technical Conference}, 2020.

\bibitem{bilal2023great}
M.~Bilal, M.~Canini, R.~Fonseca, and R.~Rodrigues, ``With great freedom comes great opportunity: Rethinking resource allocation for serverless functions,'' in \emph{Proceedings of European Conference on Computer Systems}, 2023, pp. 381--397.

\bibitem{mahgoub2022orion}
A.~Mahgoub, E.~B. Yi, K.~Shankar, S.~Elnikety, S.~Chaterji, and S.~Bagchi, ``Orion and the three rights: Sizing, bundling, and prewarming for serverless dags,'' in \emph{Proceedings of Symposium on Network System Design and Implementation}, 2022.

\bibitem{suresh2020ensure}
A.~Suresh, G.~Somashekar, A.~Varadarajan, V.~R. Kakarla, H.~Upadhyay, and A.~Gandhi, ``Ensure: Efficient scheduling and autonomous resource management in serverless environments,'' in \emph{IEEE International Conference on Autonomic Computing and Self-Organizing Systems (ACSOS)}.\hskip 1em plus 0.5em minus 0.4em\relax IEEE, 2020.

\bibitem{qiu2022reinforcement}
H.~Qiu, W.~Mao, A.~Patke, C.~Wang, H.~Franke, Z.~T. Kalbarczyk, T.~Ba{\c{s}}ar, and R.~K. Iyer, ``Reinforcement learning for resource management in multi-tenant serverless platforms,'' in \emph{Proceedings of European Workshop on Machine Learning and Systems}, 2022.

\bibitem{akhtar2020cose}
N.~Akhtar, A.~Raza, V.~Ishakian, and I.~Matta, ``Cose: Configuring serverless functions using statistical learning,'' in \emph{IEEE International Conference on Computer Communications}.\hskip 1em plus 0.5em minus 0.4em\relax IEEE, 2020, pp. 129--138.

\bibitem{wen2022stepconf}
Z.~Wen, Y.~Wang, and F.~Liu, ``Stepconf: Slo-aware dynamic resource configuration for serverless function workflows,'' in \emph{IEEE International Conference on Computer Communications}.\hskip 1em plus 0.5em minus 0.4em\relax IEEE, 2022, pp. 1868--1877.

\bibitem{zubko2022maff}
T.~Zubko, A.~Jindal, M.~Chadha, and M.~Gerndt, ``Maff: Self-adaptive memory optimization for serverless functions,'' in \emph{European Conference on Service-Oriented and Cloud Computing}.\hskip 1em plus 0.5em minus 0.4em\relax Springer, 2022, pp. 137--154.

\bibitem{eismann2021sizeless}
S.~Eismann, L.~Bui, J.~Grohmann, C.~Abad, N.~Herbst, and S.~Kounev, ``Sizeless: Predicting the optimal size of serverless functions,'' in \emph{Proceedings of International Middleware Conference}, 2021.

\bibitem{yu2023faasdeliver}
G.~Yu, P.~Chen, Z.~Zheng, J.~Zhang, X.~Li, and Z.~He, ``Faasdeliver: Cost-efficient and qos-aware function delivery in computing continuum,'' \emph{IEEE Transactions on Services Computing}, vol.~16, no.~5, pp. 3332--3347, 2023.

\bibitem{li2022faasflow}
Z.~Li, Y.~Liu, L.~Guo, Q.~Chen, J.~Cheng, W.~Zheng, and M.~Guo, ``Faasflow: Enable efficient workflow execution for function-as-a-service,'' in \emph{Proceedings of International Conference on Architectural Support for Programming Languages and Operating Systems}, 2022, pp. 782--796.

\bibitem{li2023dataflower}
Z.~Li, C.~Xu, Q.~Chen, J.~Zhao, C.~Chen, and M.~Guo, ``Dataflower: Exploiting the data-flow paradigm for serverless workflow orchestration,'' in \emph{Proceedings of International Conference on Architectural Support for Programming Languages and Operating Systems}, 2023, pp. 57--72.

\bibitem{wang2024probabilistic}
W.~Wang, Q.~Wu, Z.~Zhang, J.~Zeng, X.~Zhang, and M.~Zhou, ``A probabilistic modeling and evolutionary optimization approach for serverless workflow configuration,'' \emph{Software: Practice and Experience}, vol.~54, no.~9, pp. 1697--1713, 2024.

\bibitem{safaryan2022slam}
G.~Safaryan, A.~Jindal, M.~Chadha, and M.~Gerndt, ``Slam: Slo-aware memory optimization for serverless applications,'' in \emph{IEEE International Conference on Cloud Computing}.\hskip 1em plus 0.5em minus 0.4em\relax IEEE, 2022, pp. 30--39.

\bibitem{raza2023configuration}
A.~Raza, N.~Akhtar, V.~Isahagian, I.~Matta, and L.~Huang, ``Configuration and placement of serverless applications using statistical learning,'' \emph{IEEE Transactions on Network and Service Management}, vol.~20, no.~2, pp. 1065--1077, 2023.

\end{thebibliography}
\end{document}